\documentclass[aps,prb,showpacs,amsmath,amssymb,twocolumn,notitlepage]{revtex4-2}

\usepackage{color}
\usepackage{amsmath}
\usepackage{graphicx}
\usepackage{multirow}
\usepackage{float}
\usepackage{bm}

\DeclareMathOperator{\sgn}{sgn} 

\begin{document}

\newcommand{\bn}{{\bm n}}
\newcommand{\bp}{{\bm p}}   
\newcommand{\br}{{\bm r}}
\newcommand{\bk}{{\bm k}}
\newcommand{\bv}{{\bm v}}
\newcommand{\brho}{{\bm{\rho}}}
\newcommand{\bj}{{\bm j}}
\newcommand{\wk}{\omega_{\bf k}}
\newcommand{\nk}{n_{\bf k}}
\newcommand{\eps}{\varepsilon}
\newcommand{\la}{\langle}
\newcommand{\ra}{\rangle}
\newcommand{\be}{\begin{equation}}
\newcommand{\ee}{\end{equation}}
\newcommand{\intl}{\int\limits_{-\infty}^{\infty}}
\newcommand{\dE}{\delta{\cal E}^{ext}}
\newcommand{\SE}{S_{\cal E}^{ext}}
\newcommand{\dsp}{\displaystyle}
\newcommand{\phit}{\varphi_{\tau}}
\newcommand{\p}{\varphi}
\newcommand{\cL}{{\cal L}}
\newcommand{\dphi}{\delta\varphi}
\newcommand{\dbj}{\delta{\bf j}}
\newcommand{\dI}{\delta I}
\newcommand{\dph}{\delta\varphi}
\newcommand{\ua}{\uparrow}
\newcommand{\da}{\downarrow}
\newcommand{\ip}{\{i_{+}\}}
\newcommand{\im}{\{i_{-}\}}
\newcommand{\nnn}[1]{ #1}   
\newcommand{\ooo}[1]{#1}    

\title{Finite-frequency response of Rasba electron gas with two-particle scattering}

\author{K. E.~Nagaev} 

\affiliation{Kotelnikov Institute of Radioengineering and Electronics, Mokhovaya 11-7, Moscow 125009, Russia}

\date{\today}

\begin{abstract}
Two-dimensional systems with Rashba spin-orbit coupling are not Galilean invariant and therefore electron--electron
collisions in them may affect the current. However when taken alone, they cannot ensure a nonzero dc resistivity, so
their effects are masked by impurity scattering. Here
we calculate the related finite-frequency response and show that the electron--electron scattering in clean Rashba 
conductors decreases the Drude weight while resulting in a finite dissipative component of the response
\ooo{outside of the Drude peak}.
\end{abstract}

\maketitle

\section{Introduction}

Electron systems with spin-orbit coupling are not Galilean-invariant and therefore the electron--electron scattering may
affect the electrical current in them. A considerable amount of recent theoretical work was related with their optical conductivity
and finite-frequency absorption. Many papers dealt with single-particle absorption due to the transitions between spin-split
subbands \cite{Magarill01,Xu03,Wang05,Gumbs05,Kushwaha06,Pletyukhov06,Maiti15}, which leads to a box-shaped contribution to 
the real part of optical conductivity
at zero temperature. A negative correction to the Drude weight due to electron--electron interaction was obtained using the
time-dependent Hartree-Fock approximation \cite{Agarwal11}. In Ref. \cite{Farid06}, the finite-temperature absorption that 
results from excitation of electron--hole pairs was calculated by treating the electron--electron interaction as perturbation to the second order. 

In this paper, we calculate the finite-frequency and finite-temperature electric response of two-dimensional (2D) Rashba 
electron gas
with electron--electron scattering beyond the perturbation theory. To this end, we use the Boltzmann equation similar to that
derived recently \cite{Nagaev20} for calculating the dc conductivity of these systems. Despite the lack of Galilean invariance,
the Rashba electron gas is translationally invariant and therefore there is a perturbation of electron distribution of a 
definite form that
is stable with respect to electron--electron collisions. For this reason, a finite dc conductivity may be ensured only in the
presence of an additional mechanism of scattering like impurities that suppress this perturbation. But in the case of an ac response, the time derivative in the Boltzmann equation plays the same role as  the additional scattering and eliminates the 
divergence of the current. We
restrict ourselves to the frequencies much lower than those related with intersubband transitions.

The paper is organized as follows. Section \ref{sec:gen} presents general equations. In Section \ref{sec:stable},  the existence
of the perturbation of electron distribution immune to the electron--electron scattering is proved for an arbitrary dispersion 
law. Section \ref{sec:calc} presents the calculations and results, and finally, Section \ref{sec:discuss} contains their 
discussion. The Appendix presents some lengthy expressions.

\begin{figure}
 \includegraphics[width=1.0\columnwidth]{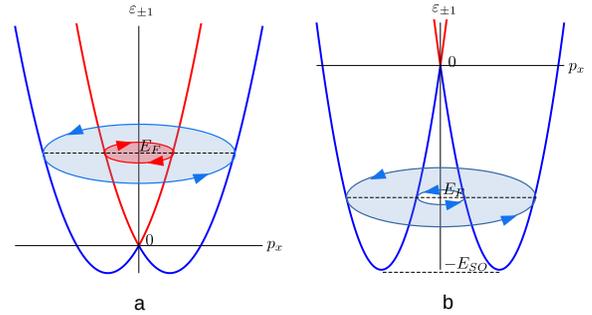}
 \caption{\label{fig:bands} Doubly connected Fermi surface (a) at $E_F>0$ and (b) at $E_F<0$. The thick red and blue
 lines show $\eps_1(p_x)$ and $\eps_{-1}(p_x)$ for $p_y=0$. Solid arrows show the directions
 of electron spin at different Fermi contours.}
\end{figure}

\section{The model} \label{sec:gen}

As a model, we consider a 2D electron gas with Rashba spin-orbit coupling (SOC), which resides in the $xy$ plane and is described
by the Hamiltonian
\be
 \hat{H} = \frac{\hat p_x^2 + \hat p_y^2}{2m} + \alpha\,(\hat\sigma_x\hat p_y - \hat\sigma_y\hat p_x),
 \label{H}
\ee
where $\alpha$ is the Rashba coupling constant and $\hat{\sigma}_{x,y}$ are the Pauli matrices. The diagonalization of this
Hamiltonian results in two subbands with dispersion laws
\be
 \eps_{\nu}(\bp) = \frac{p_x^2 + p_y^2}{2m} + \nu\alpha\sqrt{p_x^2 + p_y^2}, \quad \nu=\pm 1,
\label{eps+-}
\ee
which correspond to the two opposite helicities of electrons (see Fig. \ref{fig:bands}). These subbands are tangent
at $p_x=p_y=0$, and the Fermi surface is doubly connected below and above $E_F=0$. The minimum of the lower subband is 
given by $-E_{SO}$, where $E_{SO}=m\alpha^2/2$. The eigenstates of the Hamiltonian Eq. \eqref{H} are two-component spinors 
with components in the $\hat s_z$ basis
\be
 \psi_{\bp\nu}(\br) 
 = \frac{1}{\sqrt{2}}\,e^{i\bp\br/\hbar} \binom{e^{i\chi_{\bp}/2}}{\nu e^{-i\chi_{\bp}/2}},
 \label{WF}
\ee
where $\chi_{\bp} = \arctan(p_x/p_y)$, so the spin is  directed perpendicularly to $\bp$ either clockwise or counterclockwise.
We assume that the electron--electron interaction is weak and does not affect the electron spectrum.

\section{Boltzmann equation and collision integral} \label{sec:stable}

The most simple and convenient way to calculate the response of weakly interacting electron system to a slowly varying electric 
field as compared with the subband separation is the Boltzmann equation
\be
 \frac{\partial f_{\nu}}{\partial t} 
 + e{\bm E}(t)\,\frac{\partial f_{\nu}}{\partial\bp}
 = I_{\nu}^{ee},
 \label{Boltz-1}
\ee
where $f_{\nu}(\bp,t)$ is the electron distribution in the basis of exact eigenstates of the Hamiltonian \eqref{H}. 
The electron--electron collision integral may be written in the form
\begin{multline}
 I_{\nu}^{ee}(\bp) = \sum_{\nu_1} \sum_{\nu_2} \sum_{\nu_3}
 \int\frac{d^2p_1}{(2\pi\hbar)^2} \int\frac{d^2p_2}{(2\pi\hbar)^2} \int d^2p_3\,
 \\ \times
 \delta(\bp + \bp_1  -  \bp_2 - \bp_3)\,\delta(\eps_{\nu} + \eps_{\nu_1} - \eps_{\nu_2} - \eps_{\nu_3})
 \\ \times
 W(\nu\bp,\nu_1\bp_1;\nu_2\bp_2,\nu_3\bp_3)   
 \\ \times
 \bigl[ (1-f)(1-f_1)\,f_2\,f_3 - f\,f_1\,(1-f_2)(1 - f_3) \bigr].
 \label{Iee-1}
\end{multline}
provided that the scattering is microscopically reversible, i.~e.  
\be
 W(\nu\bp,\nu_1\bp_1;\nu_2\bp_2,\nu_3\bp_3) = W(\nu_2\bp_2,\nu_3\bp_3;\nu\bp,\nu_1\bp_1).
 \label{balance}
\ee
This equality may not hold if the system lacks either time-reversal or inversion symmetry, but its violations 
show up only beyond the Born approximation \cite{Belinicher80,Gorban14}.

Regardless of the number of subbands and the explicit form of $\eps_{\nu}(\bp)$, the collision integral Eq. 
\eqref{Iee-1}  is always turned into zero by a distribution of the form \cite{Maslov11}
\be
 f_{\nu}(\bp) = \bar{f}(\eps_{\nu} + {\bm u}\bp),
 \label{u}
\ee
where $\bm u$ is an arbitrary constant vector and $\bar{f}(\eps_{\nu})$ is the Fermi distribution function. 
With this substitution, one easily obtains that
\begin{multline}
 (1-f)(1-f_1)\,f_2\,f_3 = f f_1 f_2 f_3
 \\ \times
 \exp\left[\frac{\eps_{\nu}(\bp) + \eps_{\nu_1}(\bp_1) + {\bm u}\,(\bp+\bp_1)}{T}\right] 
\end{multline}
and a similar equality for $f f_1\,(1-f_2)(1 - f_3)$.
Therefore the difference in the square brackets in Eq. \eqref{Iee-1} turns into zero because of the momentum
and energy conservation. As a consequence, 
\nnn{any perturbation} of the form
\be
 \delta f_{\nu}(\bp) = \bm u\bp\,\bar{f}(\eps_{\nu})\,[1 - \bar{f}(\eps_{\nu})]
 \label{df_u}
\ee
\nnn{turns the collision integral into zero to the first approximation}. 
This is the reason why the electron--electron scattering alone cannot ensure a finite
dc conductivity even for multiband electron systems \cite{Nagaev20}.

\section{Electrical response} \label{sec:calc}

We assume that the electron--electron interaction \nnn{in the conductor is screened by a nearby metallic gate such
that the distance to the gate $d_0$ is smaller than the Fermi wavelength,}
\ooo{so that the interaction potential may be presented in the form $V(\br-\br') = 4\pi e^2d_0\,\kappa^{-1}\,\delta(\br-\br')$,
where $e$ is the electron charge and $\kappa$ is the dielectric constant.}
Therefore the scattering probability $W$ in Eq. \eqref{Iee-1} may be calculated in the Born approximation
\nnn{and equals
\begin{multline}
 W(\nu\bp,\nu_1\bp_1;\nu_2\bp_2,\nu_3\bp_3)
 = 8\pi^3\,\frac{e^4 d_0^2}{\hbar\kappa^2}\,
 [1 - \nu\,\nu_1 \cos(\widehat{\bp,\bp_1})]
 \\ \times
 [1 - \nu_2\,\nu_3 \cos(\widehat{\bp_2,\bp_3})].
 \label{W4}
\end{multline}
}
As the system is rotationally symmetric in the $xy$ plane, 
the linear response to the electric field $\bm E(t)$ may be conveniently sought in the form
\be
 f_{\mu}(\eps,\p) = \bar f(\eps) + C_{\mu}(\eps)\,\bar f(\eps)\,[1 - \bar f(\eps)]\cos\p,
 \label{f-ansatz}
\ee
where $\mu=\pm 1$ labels the Fermi contours with larger momentum $p_1$ and smaller momentum  $p_{-1}$ for a given $E_F$, 
the energy $\eps$ is measured from $E_F$, 
and $\p$ is the angle between $\bm E$ and $\bp$. 
Note that the corresponding velocities $v_{\mu}=(\partial p_{\mu}/\partial\eps)^{-1}$ for the same $\eps$ are equal  
at $E_F>0$ 
but have opposite signs at $E_F<0$, so that $v_{-1}= v_1 \sgn E_F$. Assuming that  $p_{\mu}$ and
$v_{\mu}$ are independent of energy near the Fermi level and that the coefficients $C_{\mu}$ are even
functions of $\eps$, one may bring the linearized collision integral Eq. \eqref{Iee-1} to the form \cite{Nagaev20}
\begin{multline}
 I_{\mu}^{ee}(\eps,\p) = \frac{2\,\Gamma_{ee}\cos\p}{T^2} \int\! d\eps' K(\eps,\eps')
 \biggl\{
  \ln\frac{E_F}{T}\,\Phi_{\mu}  
  \bigl[ C_{\mu}(\eps') \\- C_{\mu}(\eps) \bigr]
  +
  \Psi_{\mu}\,
  \frac{p_{\mu}\,C_{-\mu}(\eps') - p_{-\mu}\,C_{\mu}(\eps')}{p_{1} + p_{-1}}
 \biggr\},
 \label{Iee-2}
\end{multline}
where \nnn{$\Gamma_{ee}(T)=16\pi^2e^4\,d_0^2\,T^2\,(p_{1}+p_{-1})/32\pi^3\hbar^5\kappa^2v_{1}^3$} is the effective 
rate of electron--electron collisions,
\begin{gather}
 K(\eps,\eps') = \bigl[1 - \bar{f}(\eps)\bigr]\,\frac{\eps-\eps'}{e^{(\eps-\eps')/T} - 1}\,\bar{f}(\eps'),
 \label{K}
\\
 \Phi_{\mu} =  4\,\frac{p_{\mu} + 3\,p_{-\mu}}{p_{1} + p_{-1}},
 \label{Phi}
\end{gather}
and the explicit expressions for the dimensionless functions $\Psi_{\mu}(E_F/E_{SO})$ are given in the Appendix. It is 
clearly seen that $I^{ee}_{\mu}(\eps,\p)$ is turned into zero by the distribution of the form Eq. \eqref{df_u}, i. e. with 
$p_1 C_{-1} = p_{-1} C_1 = {\rm const}(\eps)$. The logarithmic singularity in the first term in curly brackets in Eq.
\eqref{Iee-2} is a characteristic feature of 2D scattering that results from head-on and small-angle collisions and manifests
itself in the inverse quasiparticle lifetime \cite{Hodges71,Giuliani82} and thermal conductivity \cite{Lyakhov03} for 2D
conductors with singly connected Fermi surface.

Assume that the electric field has a sinusoidal time dependence ${\bm E}(t) = e^{-i\ooo{(\omega+i\delta)} t}\,{\bm E}_{\omega}$,
\ooo{where $\delta$ is infinitely small  and positive.}
With the substitution Eq. \eqref{f-ansatz}, the Boltzmann equation Eq. \eqref{Boltz-1} results in a system of two integral 
equations in $C_{\mu}(\eps)$. This system may be solved by the method first proposed by Brooker and Sykes \cite{Brooker68}. To 
this end, we introduce a new variable
\be
\rho_{\mu}(\eps) = [\bar{f}\,(1-\bar{f})]^{1/2}\,C_{\mu}(\eps),
\label{rho}
\ee
which makes the kernel of the integral in Eq. \eqref{Iee-2} a function of the difference $\eps'-\eps$. Therefore the system
of integral equations may be brought to the differential form by a Fourier transform
\be
 \tilde\rho_{\mu}(u) = \int d\eps\,e^{-i\eps u}\,\rho_{\mu}(\eps).
 \label{Fourier}
\ee
A subsequent introduction of the new independent variable $\xi=\tanh(\pi Tu)$ brings Eqs. \eqref{Boltz-1} to the form
\begin{multline}
 \Gamma_{ee} \left[
  \ln\frac{E_F}{T}\,\Phi_{\mu} \left(\hat{L} + 2\right)\,\tilde\rho_{\mu}
 -
 2\,\Psi_{\mu}\,\frac{p_{-\mu}\,\tilde\rho_{\mu} - p_{\mu}\,\tilde\rho_{-\mu}}
                                {p_{1} + p_{-1}}
 \right]
\\ +
 \frac{1}{\pi^2}\,\frac{i\omega}{1-\xi^2}\,\tilde\rho_{\mu}
 =
 -\frac{1}{\pi}\,\frac{eEv_{\mu}}{\sqrt{1 - \xi^2}},     
 \label{Boltz-xi}                            
\end{multline}
where $\hat L$ is the differential operator
\be
 \hat{L}\,\phi = \frac{\partial}{\partial\xi}\biggl[(1-\xi^2)\,\frac{\partial\phi}{\partial\xi}\biggr]
 - \frac{\phi}{1-\xi^2}.
 \label{L}
\ee
The solutions of Eqs. \eqref{Boltz-xi} may be presented in the form of a series
\be
 \tilde\rho_{\mu}(\xi) = \sum_{m=0}^{\infty} \gamma_{\mu m}\,\phi_{2m}(\xi),
 \label{series}
\ee
where $\phi_m(\xi)$ are the eigenfunctions of operator $\hat L$ with corresponding eigenvalues $-(m+1)(m+2)$
\cite{Landau-book}. A substitution of the expansions Eqs. \eqref{series} into Eqs. \eqref{Boltz-xi} results 
in an infinite system of equations for the coefficients $\gamma_{\mu m}$ of the form
\begin{multline}
 2\,\Gamma_{ee} \biggl[
  m\,(2m+3)\,\ln\frac{E_F}{T}\,\Phi_{\mu}\,\gamma_{\mu m} 
\\  + 
  \Psi_{\mu}\,\frac{p_{-\mu}\,\gamma_{\mu m} - p_{\mu}\,\gamma_{-\mu m}}{p_{1} + p_{-1}}
 \biggr]
-
 i\,\frac{\omega}{\pi^2} \sum_{n=0}^{\infty} Y_{mn}\,\gamma_{\mu n}
\\ =
 eEv_{\mu}\,X_m/\pi.
 \label{Boltz-gamma}
\end{multline}
where $Y_{mn}$ are the matrix elements of $(1-\xi^2)^{-1}$ between $\phi_{2m}$ and $\phi_{2n}$, and $X_m$ are the projections
of $(1-\xi^2)^{-1/2}$ on $\phi_{2m}$. The explicit expressions for these quantities are given in the Appendix. 
\nnn{The current density is given by the sum} \ooo{\cite{Nagaev20}} \nnn{
\be
 j = \frac{e}{8\pi^2\hbar^2} \sum_{\mu} p_{\mu} \sgn v_{\mu}
 \sum_m X_m\,\gamma_{\mu m}.
 \label{j-series}
\ee
}

In the high-frequency limit, one easily obtains directly from Eq. \eqref{Boltz-1} that $C_{\mu} = eE_{\omega} v_{\mu}/\omega T$ 
and therefore the imaginary part of the response is
\be
 \sigma''_0 = \frac{e^2}{4\pi\hbar^2\omega}\,v_1(p_1 + p_{-1}),
 \label{sigma0}
\ee
while the dissipative part $\sigma'$ is zero. Equation \eqref{sigma0} coincides with the results of \cite{Maiti15}.
In the opposite limit $\Gamma_{ee} \gg \omega$, one cannot simply set $\omega=0$ 
because of the existence of the perturbation \eqref{df_u} with zero relaxation rate, which leads to the divergence of
$\gamma_{\mu0}$. This divergence can be eliminated by keeping $\omega$ small but finite and isolating the most singular
in $\omega$ contribution to $\sigma$. This contribution may be obtained by setting $\gamma_{\mu m}=0$ for all $m \ne 0$
and solving the resulting systems of two equations for $\gamma_{\mu 0}$. The resulting conductivity \ooo{is}
\begin{multline}
 \ooo{\sigma_{ee}} = \ooo{i}\,\frac{e^2}{4\pi\hbar^2}\,\frac{v_1}{\ooo{\omega+i\delta}}
 \\ \times
 \frac{(p_1^2 + \sgn E_F\,p_{-1}^2)(\Psi_{-1} + \sgn E_F\,\Psi_1)}{\Psi_{-1}\,p_1 + \Psi_{1}\,p_{-1}}.
 \label{sigma''}
\end{multline}
\ooo{At $\omega\ne 0$, it}
is purely imaginary and inversely proportional to $\omega$, like $\sigma_0''$. Though it is temperature-independent,
it still depends on the properties of electron--electron scattering through the quantities $\Psi_{\pm 1}$.

\ooo{In the leading approximation, the real part of conductivity is proportional to $\delta(\omega)$.}
\nnn{The dissipative part of conductivity \ooo{at $\omega\ne 0$} is given by the subleading term, \ooo{which is}  
independent of $\omega$. To calculate it, 
one should, in principle, take into account the components of Eq. \eqref{series} with higher $m$. The most singular 
parts of $\gamma_{\mu 0}$ should be substituted into Eqs. \eqref{Boltz-gamma} with $m \ne 0$, and the solutions should 
be substituted back into Eqs. \eqref{Boltz-gamma} with $m=0$. However because of the condition $\ln(E_F/T)\gg 1$,
the contribution from $\gamma_{\mu m }$ with $m>0$ is small, and it is sufficient to find the subleading term in the
equations with $m=0$. Therefore} \ooo{at $\omega\ne 0$,}
\begin{multline}
 \sigma_{ee}' = \frac{3}{16\pi^3}\,\frac{e^2}{\hbar^2}\,
 \frac{v_1\,(p_1 + p_{-1})}{\Gamma_{ee}}
 \\ \times
 \frac{(p_1 - \sgn E_F\,p_{-1})(\Psi_{-1} p_{-1} - \sgn E_F\,\Psi_1 p_1)}
      {(\Psi_{-1} p_1 + \Psi_1 p_{-1})^2}.
 \label{sigma'}    
\end{multline}
It is noteworthy that neither $\sigma'_{ee}$ nor $\sigma''_{ee}$ contains the $\ln^{-1}(E_F/T)$ factor,
much like the dc conductivity in a presence of weak impurity scattering \cite{Nagaev20}.

\nnn{At $E_{SO}\ll E_F$, the scattering corrections to the conductivity are proportional to $E_{SO}^{3/2}$, so that
\be
 \sigma'_{ee} \approx \frac{3\,e^2E_{SO}^{3/2}}{64\pi^3\hbar^2\Gamma_{ee}E_F^{1/2}},
 \quad
 \sigma''_{ee}-\sigma''_0 \approx -\frac{e^2 E_{SO}^{3/2}}{\pi\hbar^2\omega E_F^{1/2}}.
\ee
}

\begin{figure}
 \includegraphics[width=0.9\columnwidth]{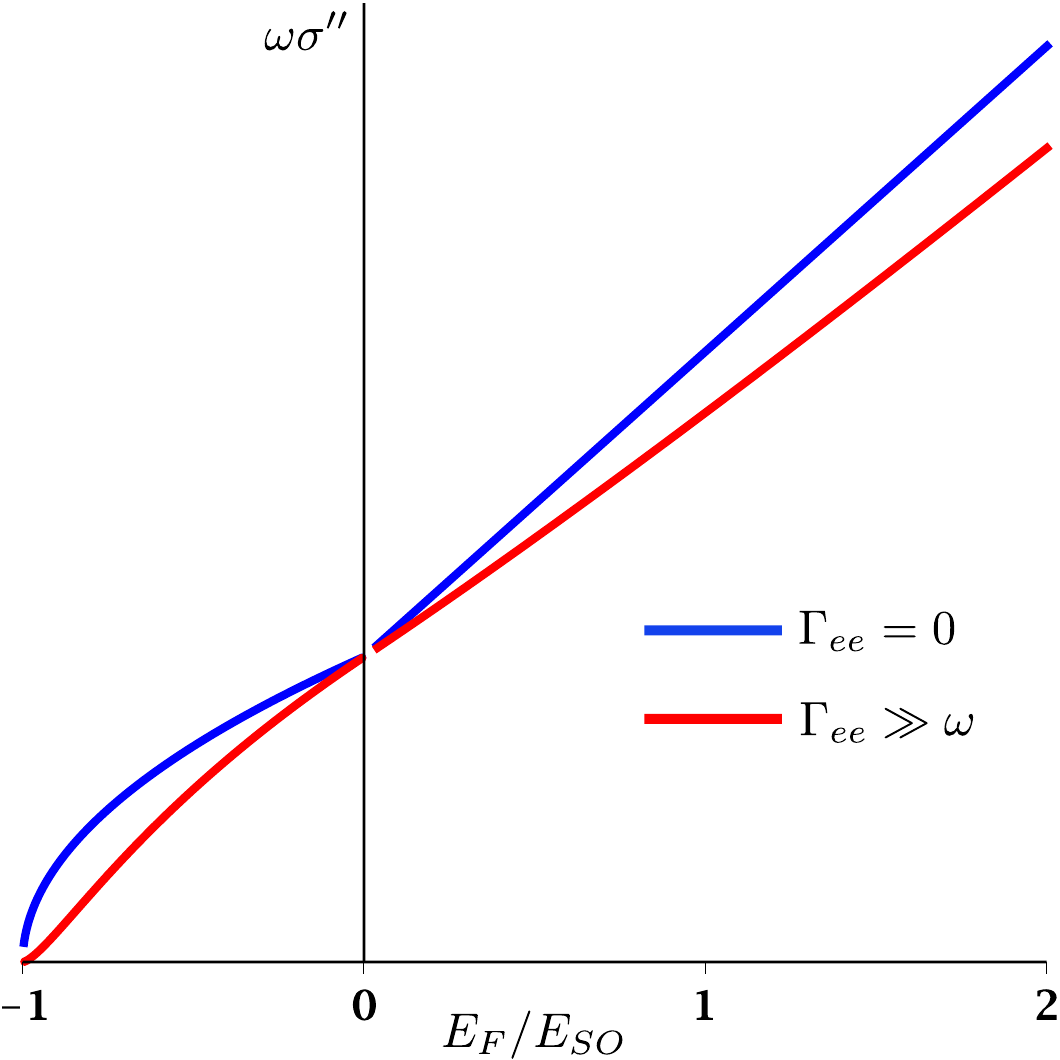}
 \caption{\label{fig:imaginary} The dependence of the imaginary part of conductivity $\sigma''$ times $\omega$ on $E_F$ 
 in the zero-temperature limit (blue curve) and in the high-temperature limit (red curve).}
\end{figure}

\begin{figure}
 \includegraphics[width=0.9\columnwidth]{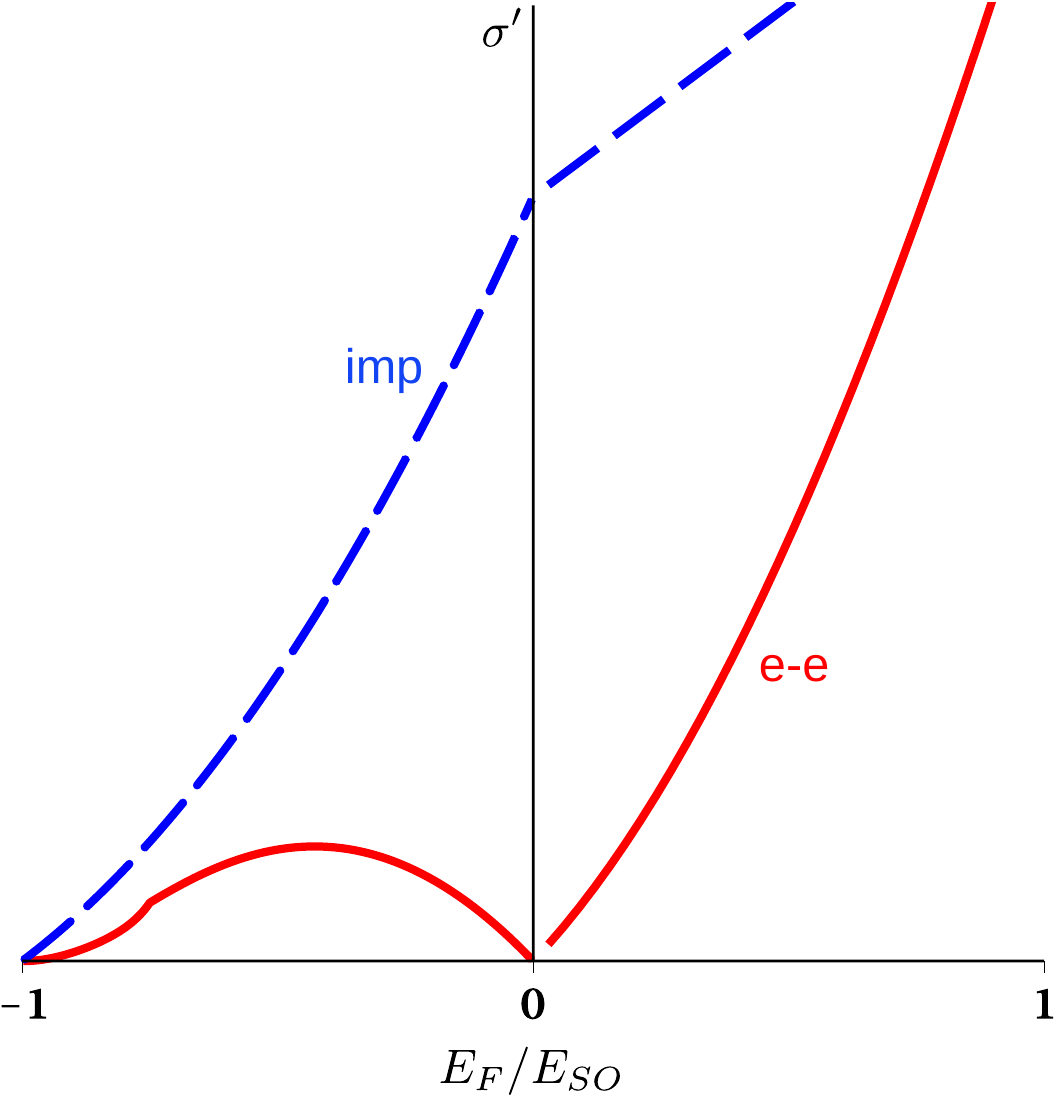}
 \caption{\label{fig:real} The dependence of the real part of conductivity $\sigma'$ on $E_F$ \ooo{at $\omega\ne0$}. The solid 
 red line shows $\sigma'_{ee}$ in the limit $\Gamma_{ee}\gg\omega$ in arbitrary units. For comparison, the dashed blue line 
 shows a sketch of $\sigma'$ in the limit of strong impurity scattering.}
\end{figure}

\section{Discussion} \label{sec:discuss}

The dependence of the imaginary and real parts of conductivity on the Fermi energy at $\omega\ll E_{SO}$ is shown in Figs.
\ref{fig:imaginary} and \ref{fig:real}. It is clearly seen that $\sigma''_{ee}$ is suppressed relatively to 
$\sigma''_0$ both at $E_F>0$ and $E_F<0$ \ooo{and} \nnn{the electron--electron scattering reduces the Drude weight}. 
This suppression is especially pronounced near the bottom of the lower subband,
where $\sigma_0'' \propto (E_F+E_{SO})^{1/2}$ and $\sigma_{ee}''\propto (E_F+E_{SO})^{3/2}$. At $E_F=0$, the 
electron--electron scattering does not affect the conductivity. It is noteworthy that
unlike $\sigma''_0$, $\sigma''_{ee}$ exhibits no kink and $d\sigma_{ee}''/dE_F$ is continuous at this point.

In contrast to the imaginary part, the dissipative part $\sigma_{ee}'$ scales as $T^{-2}$ even at $\Gamma_{ee} \gg\omega$.
Another clear distinction from $\sigma_{ee}''$ is the  nonmonotone $\sigma_{ee}'(E_F)$ dependence with $\sigma_{ee}'(0)=0$.
The disappearance  of dissipation at $E_F=0$ is quite natural because the inner Fermi contour shrinks into a point at this
Fermi energy and the Fermi surface becomes effectively singly connected. Our results for $\sigma'$ sharply differ from the 
$T^2/\omega^2$ dependence obtained in \cite{Farid06} in the low-frequency limit using second-order perturbation theory in 
$V_0$. The suppression of $\sigma''$ is a consequence of the emergence of a finite dissipation, in agreement with the 
Kramers--Kronig relations. Unlike the suppression predicted by Agarwal et al. \cite{Agarwal11}, it is temperature-dependent
and takes place even at weak electron--electron interaction.

\ooo{Suppose now that the material is not perfectly clean.}
\nnn{An important question is how the dissipative conductivity $\sigma''_{ee}\propto\Gamma_{ee}^{-1}$ given by 
Eq. \eqref{sigma'} is related with the dc conductivity found  to be inversely proportional to the 
impurity scattering rate $\Gamma_{imp}$ in Ref. \cite{Nagaev20}.  The answer is given by Fig. \ref{fig:combined}, 
which shows the frequency dependence 
of $\sigma'$ in a presence of the electron--electron and much weaker impurity scattering.  This dependence
exhibits two clearly seen plateaus at $\omega \le \Gamma_{imp}$ and at $\omega \gg \Gamma_{imp}$. The former 
plateau presents the results of \cite{Nagaev20}, while the latter corresponds to Eq. \eqref{sigma'}. In other words, the
results of \cite{Nagaev20} correspond to the limit $\omega \ll \Gamma_{imp}$, while Eq. \eqref{sigma'} corresponds to
$\omega\gg\Gamma_{imp}$. \ooo{Naturally, the presence of impurities eliminates the $\delta(\omega)$ peak.}
\begin{figure}
 \includegraphics[width=0.9\columnwidth]{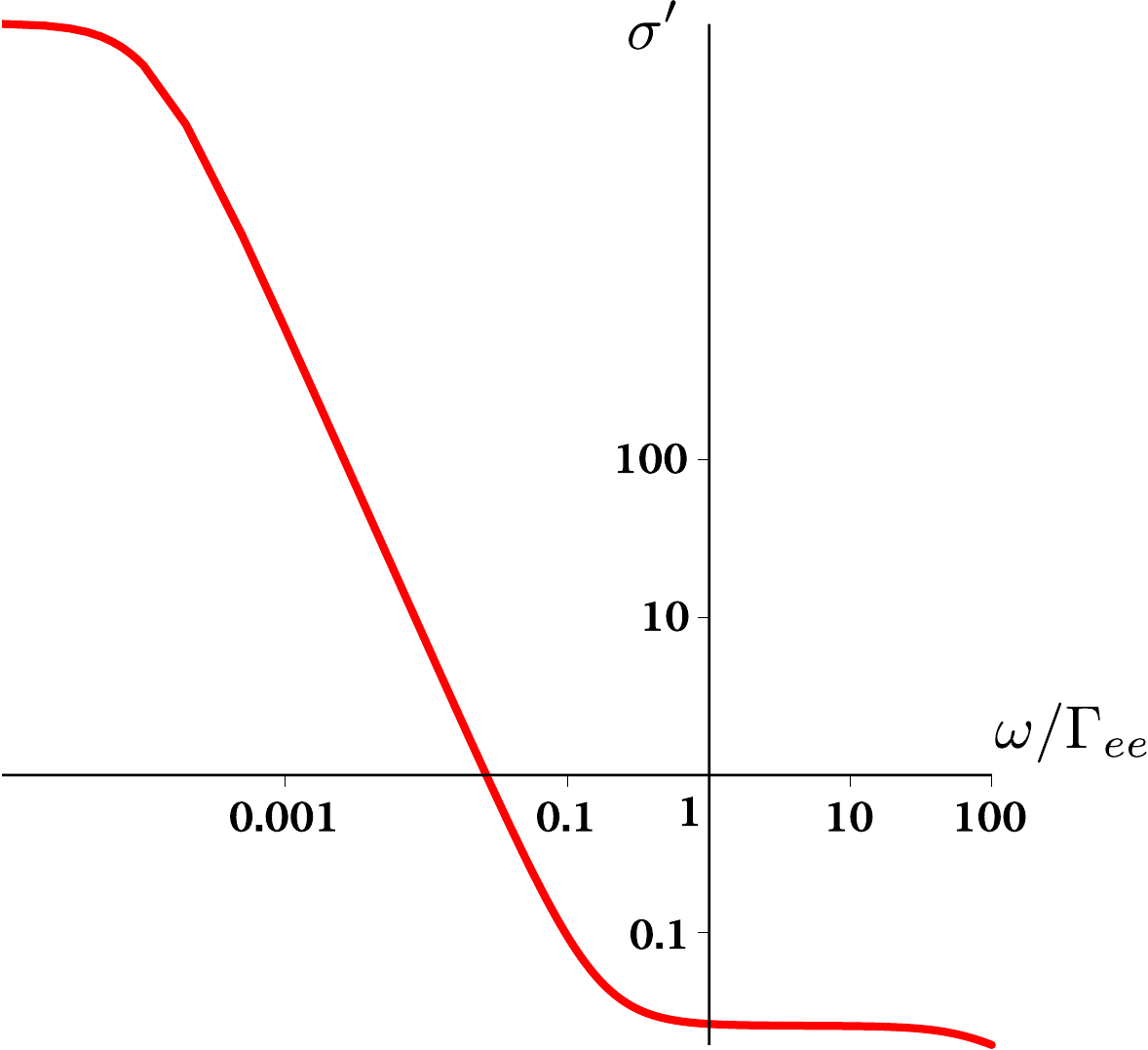}
 \caption{\label{fig:combined} A log-log plot of $\sigma'$ in arbitrary units vs $\omega/\Gamma_{ee}$ for $E_F/E_{SO}=-0.85$ 
 and $\Gamma_{imp} = 10^{-4}\,\Gamma_{ee}$.}
\end{figure}
}

The finite-frequency dissipation may be more convenient for experimental investigations of electron--electron scattering
than the dc conductivity because it does not depend on the type of impurities in the conductor
\footnote{This dependence was noted in \cite{Nagaev20}}.
Though this dissipation is nonzero both above and below the band-crossing point $E_F=0$, it may be more 
conveniently observed below this point. One of the conditions is $\omega\ll\Gamma_{ee}$. 
For example, for InAs, which exhibits 
Rashba parameter $\hbar\alpha =1.2\,{\rm eV\AA}$ \cite{Heedt2017}, the electron concentration $10^{10}$ cm$^{-2}$, 
the 2D gas--gate distance of 100 nm, and $T=4$ K, the frequency has to be smaller than $\Gamma_{ee} \approx 0.5$ THz. 
On the other hand, these experiments would require high-quality samples because the electron--impurity scattering length has 
to be much larger than $l_{ee} \sim 40$ nm.

\nnn{The above calculations were performed for point-like electron--electron interactions because this model allows an
analytical solution. However the existence of zero-relaxation modes, which is their cornerstone, stems from very general 
properties of the system like translational invariance and Fermi statistics, so the suppression of Drude weight and
finite dissipation are not the consequences  of point-like interaction and  should be also observed for a long-range potential.}

\begin{acknowledgements}

This work was carried out within the framework of the state task.

\end{acknowledgements}

\appendix*
\section{Explicit expressions for some quantities}

The quantities $\Phi_{\mu}$ and $\Psi_{\mu}$ in the collision integral Eq. \eqref{Iee-2} are defined by the expressions
\begin{multline}
\ln\frac{E_F}{T}\,
 \Phi_{\mu} = \sum_{\mu_1} \sum_{\mu_2} \sum_{\mu_3} \frac{p_{\mu_1}}{p_{\mu} +p_{-\mu}}
 \int_{-\pi}^{\pi} d\chi\,(1 - \mu\,\mu_1\cos\chi)
\\ \times
 \Theta({\cal D}_{\mu..\mu_3})\,
 {\cal D}_{\mu..\mu_3}^{\mu_2\mu_3/2},
 \label{Q-def}
\end{multline}
and
\begin{multline}
 \Psi_{\mu} = \sum_{\mu_1} \sum_{\mu_2} \sum_{\mu_3} \frac{p_{\mu_1}}{p_{-\mu}}
 \int_{-\pi}^{\pi} d\chi\,(1 - \mu\,\mu_1\cos\chi)\,
 \Theta({\cal D}_{\mu..\mu_3})
\\ \times
 {\cal D}_{\mu..\mu_3}^{\mu_2\mu_3/2}\, 
 \left( 1 - \delta_{\mu\mu_1}\cos\chi - 2\,\delta_{\mu\mu_2}\lambda_{\mu..\mu_3} \right),
 \label{Psi-def}
\end{multline}
where
 $\mu..\mu_3$ stands for $\mu\mu_1\mu_2\mu_3$,
\be
 {\cal D}_{\mu..\mu_3} 
 = \frac{(p_{2} + p_{3})^2 - p^2 - p_{1}^2 - 2\,p\,p_{1}\cos\chi}
        {p_{}^2 + p_{1}^2 + 2\,p\,p_{1}\cos\chi - (p_{2} - p_{3})^2},
 \label{D}
\ee
and
\begin{multline}
 \lambda_{\mu..\mu_3}
 = 
 \frac{1}{2}\,
 \frac{p_{2}^2 - p_{3}^2 + p_{}^2 + p_{1}^2 + 2\,p_{}\,p_{1}\cos\chi}
      {p_{}^2 + p_{1}^2 + 2\,p_{}\,p_{1}\cos\chi}
\\ \times      
 (p_{} + p_{1}\cos\chi)/{p_{2}}.
\end{multline}
%

The normalized eigenfunctions of differential operator $\hat L$ Eq. \eqref{L} are given by 
the expressions
\be
 \phi_m(\xi) = \sqrt{\frac{(2m+3)(m+2)}{8\,(m+1)}} \sqrt{1-\xi^2}\,P_m^{(1,1)}(\xi),
 \label{psi_m}
\ee
where $P_m^{(1,1)}(\xi)$ are Jacobi polynomials. The coefficients of expansion of $(1-\xi^2)^{-1/2}$
in these functions are given by 
\be
 X_m = \int_{-1}^1 d\xi\, \frac{\phi_{2m}(\xi)}{\sqrt{1-\xi^2}}
 =
 \sqrt{\frac{4m+3}{(2m+1)(m+1)}}.
 \label{X}
\ee
The matrix elements of $1/(1-\xi^2)$ between the eigenfunctions of $\hat L$ are given by the equation
\begin{multline}
 Y_{mn} = \int_{-1}^1 d\xi\,\frac{\phi_{2m}(\xi)\,\phi_{2n}(\xi)}{1 - \xi^2}
 =
 \frac{\min(m,n) + 1/2}{\max(m,n) + 1}
\\ \times
 \sqrt{\frac{ (4m+3)(m+1)(4n+3)(n+1) }{ (2m+1)(2n+1) }}.
 \label{Y}
\end{multline}

\bibliography{ac,SOI-2D,ee,books}

\end{document}